\documentclass[a4paper]{article}

\usepackage{emi}
\usepackage[english]{babel}

\title{Cosmic--Ray Positron Identification with the PAMELA experiment}

\shorttitle{The PAMELA CR e$^+$ Spectrum}

\authors{
O.~Adriani$^{1,2}$, 
G.C.~Barbarino$^{3,4}$, 
G.A.~Bazilevskaya$^{5}$, 
R.~Bellotti$^{6,7}$, 
A.~Bianco$^{8}$,
M.~Boezio$^{8}$, 
E.A.~Bogomolov$^{9}$, 
M.~Bongi$^{1,2}$, 
V.~Bonvicini$^{8}$, 
S.~Bottai$^{2}$, 
A.~Bruno$^{7}$, 
F.~Cafagna$^{7}$, 
D.~Campana$^{4}$,
R.~Carbone$^{8}$, 
P.~Carlson$^{10}$, 
M.~Casolino$^{11,12}$, 
G.~Castellini$^{13}$,
C.~De~Donato$^{11}$,
M.P.~De~Pascale$^{11,14,\dag}$, 
C.~De~Santis$^{11,14}$,
N.~De~Simone$^{11}$, 
V.~Di~Felice$^{11}$, 
V.~Formato$^{8,15}$, 
A.M.~Galper$^{16}$,
A.V.~Karelin$^{16}$,
S.V.~Koldashov$^{16}$,
S.~Koldobskiy$^{16}$,
S.Yu.~Krut'kov$^{9}$, 
A.N.~Kvashnin$^{5}$, 
A.~Leonov$^{16}$,
V.~Malakhov$^{16}$,
L.~Marcelli$^{14}$,
M.~Martucci$^{14,17}$
A.G.~Mayorov$^{16}$,
W.~Menn$^{18}$, 
M.~Merg\`{e}$^{11,14}$
V.V.~Mikhailov$^{16}$,
E.~Mocchiutti$^{8}$, 
A.~Monaco$^{7}$, 
N.~Mori$^{2}$,
R.~Munini$^{8,15}$, 
G.~Osteria$^{4}$,
F.~Palma$^{11,14}$,
P.~Papini$^{2}$, 
M.~Pearce$^{10}$, 
P.~Picozza$^{11,14}$, 
C.~Pizzolotto$^{19,20,*}$, 
M.~Ricci$^{17}$, 
S.B.~Ricciarini$^{2}$,
L.~Rossetto$^{10}$, 
R.~Sarkar$^{8}$, 
M.~Simon$^{18}$, 
V.~Scotti$^{3,4}$,
R.~Sparvoli$^{11,14}$, 
P.~Spillantini$^{1,2}$, 
S.J.~Stochaj$^8$,
J.C.~Stockton$^8$,
Y.I.~Stozhkov$^{5}$, 
A.~Vacchi$^{8}$, 
E.~Vannuccini$^{2}$,
G.~Vasilyev$^{9}$, 
S.A.~Voronov$^{16}$,
Y.T.~Yurkin$^{16}$,
G.~Zampa$^{8}$, 
N.~Zampa$^{8}$,
V.G.~Zverev$^{16}$
}

\afiliations{
$^{1}$ University of Florence, Department of Physics, I-50019 Sesto Fiorentino, Florence, Italy \\
$^{2}$ INFN, Sezione di Florence, I-50019 Sesto Fiorentino, Florence, Italy \\
$^{3}$ University of Naples ``Federico II'', Department of Physics, I-80126 Naples, Italy \\
$^{4}$ INFN, Sezione di Naples, I-80126 Naples, Italy \\
$^{5}$ Lebedev Physical Institute, RU-119991, Moscow, Russia \\
$^{6}$ University of Bari, Department of Physics, I-70126 Bari, Italy \\
$^{7}$ INFN, Sezione di Bari, I-70126 Bari, Italy \\
$^{8}$ INFN, Sezione di Trieste, I-34149 Trieste, Italy \\
$^{9}$ Ioffe Physical Technical Institute, RU-194021 St. Petersburg, Russia \\
$^{10}$ KTH, Department of Physics, and the Oskar Klein Centre for Cosmoparticle Physics, AlbaNova University Centre, SE-10691 Stockholm, Sweden \\
$^{11}$ INFN, Sezione di Rome ``Tor Vergata'', I-00133 Rome, Italy \\
$^{12}$ RIKEN, Advanced Science Institute, Wako-shi, Saitama, Japan\\
$^{13}$ IFAC, I-50019 Sesto Fiorentino, Florence, Italy \\
$^{14}$ University of Rome ``Tor Vergata'', Department of Physics, I-00133 Rome, Italy \\
$^{15}$ University of Trieste, Department of Physics, I-34147 Trieste, Italy \\
$^{16}$ NRNU MEPhI, RU-115409 Moscow, Russia \\
$^{17}$ INFN, Laboratori Nazionali di Frascati, Via Enrico Fermi 40, I-00044 Frascati, Italy \\
$^{18}$ Universit\"at Siegen, Department of Physics, D-57068 Siegen, Germany \\
$^{19}$ INFN, Sezione di Perugia, I-06123 Perugia, Italy \\
$^{20}$ Agenzia Spaziale Italiana (ASI) Science Data Center, I-00044 Frascati, Italy \\

\par\noindent
$^{(*)}$ Previously at INFN, Sezione di Trieste, I-34149 Trieste, Italy \\
$^{(\dag)}$ Deceased \\
}

\email{Emiliano.Mocchiutti@ts.infn.it}

\abstract{
The PAMELA satellite borne experiment is designed to study cosmic 
rays with great accuracy in a wide energy range. One of PAMELA's main 
goal is the study of the antimatter component of cosmic rays.
The experiment, housed on board the Russian satellite Resurs--DK1, was 
launched on June 15th 2006 and it is still taking data. In this work we 
present the measurement of galactic positron energy spectrum in the 
energy range between 500 MeV and few hundred GeV.
}

\begin{document}
\maketitle

%Begin a section.
\section{Introduction}
PAMELA is a dedicated satellite borne experiment conceived by the WiZard collaboration to study the anti--particle component of the cosmic radiation. The instrument is installed 
inside a pressurized container attached to the Russian Resurs--DK1 satellite that was launched into Earth orbit by a Soyuz--U rocket on June 15$^{\mbox{th}}$ 2006 from the Baikonur 
cosmodrome in Kazakhstan. 

In this work we describe the procedure used to obtained the PAMELA results in the measurement of galactic positron flux after seven years of data taking.

\section{PAMELA}
PAMELA is build around a permanent magnet spectrometer. Six planes of double-sided silicon detectors precisely measure the deflection, and hence the rigidity and the energy, of 
charged particles in the magnetic field. A time of flight system provides the trigger to the whole apparatus, measures the particle beta and, by ionization losses, the absolute value 
of the particle charge. A calorimeter and a neutron detector are used for particle identification. An anticounter system permit to reject, in the offline analysis, the events out of 
the apparatus acceptance.

Detailed technical informations about the PAMELA instrument and launch preparations can be found in~\cite{pamelone}.

\section{Positron selection}
Protons are the main source of background in the positron sample and an excellent positron identification is needed to reduce the contamination at a negligible level.

The proton background estimation method was used to obtain the published results~\cite{psarticle,psarticle2}.
This approach consists in keeping a high selection efficiency and in quantifying the residual proton contamination by the mean of a so--called ``spectral analysis''. 
The proton distributions needed to estimate the contamination are obtained in a conservative approach using the flight calorimeter data without any dependence on simulations or test beam 
data (so called ``pre--sampler'' method). This conservative approach required strong restrictions on the calorimeter acceptance~\cite{boe09,moc11}, hence reducing 
significantly the available statistics.

In order to fully exploit the PAMELA data a new approach has been developed. In this case the full calorimeter is used for the identification. Since the ``pre--sampler'' method is not used, it is 
not possible to select a proton sample from flight data in order to tune and test the data selection criteria. A GEANT4 simulation is used instead for studying proton rejection and 
for background estimation. A neural network approach permits, furthermore, a higher selection efficiency to be achieved.

Depending on the particle energy, three different methodologies are used:
\begin{itemize}
\item neural network spectral analysis: low energies, between 1.5 and 20 GeV;
\item third momentum weighted average: higher energies, between 20 and 200 GeV;
\item lower limit estimation: highest energies, between 200 and 300 GeV.
\end{itemize}
In any case a basic and high efficient positron pre-selection is applied in order to have a reliable track reconstruction, to clean the data sample and to reduce the proton 
contamination. Positron energy is determined by the tracker deflection measurement and charge sign selection is obtained combining deflection and time of flight measurements.

\subsection{Neural networks spectral analysis}
In the energy range 1.5 - 20 GeV, positron identification against proton background can be easily achieved using the calorimeter and the energy-momentum match. However, due to PAMELA 
satellite orbit, charged particles at these rigidities, are modulated by the Earth's magnetic field. At a given position above the Earth surface only particle with 
rigidity greater than the geomagnetic cutoff can reach the apparatus. Moreover, leptons can undergo Bremsstrahlung losses in the upper part of the detectors arriving in the 
spectrometer with a smaller energy respect to their original one. Hence, galactic cosmic ray positrons can be selected only after accounting for energy losses in the apparatus. This 
is done statistically on the overall flux with a flux unfolding procedure.

To obtain the positron flux, the following steps are applied: events are grouped according to their rigidity, as measured by the spectrometer, and to their cutoff, 
determined by the satellite position using the St\"ormer vertical cutoff approximation. For each of these sets of events a Multi-Layer Perceptron (MLP) neural network algorithm is used event by event basis. 
The MLP has been configured using 24 input calorimeter observables, three hidden neurons and 900 epochs. Observables represent the shower developement in the calorimeter, taking into account transveral 
and longitudinal shower profile, hit multiplicity at the first and last calorimeter layers, energy release along the track and evolution of the shower compared to the theoretical expectation for an 
electromagnetic shower with starting energy as the one measured by the tracking system. A sample of simulated electrons and protons is randomly splitted in two halves which are used separately for 
training and testing.
 \begin{figure}[t]
  \centering
  \includegraphics[width=0.4\textwidth]{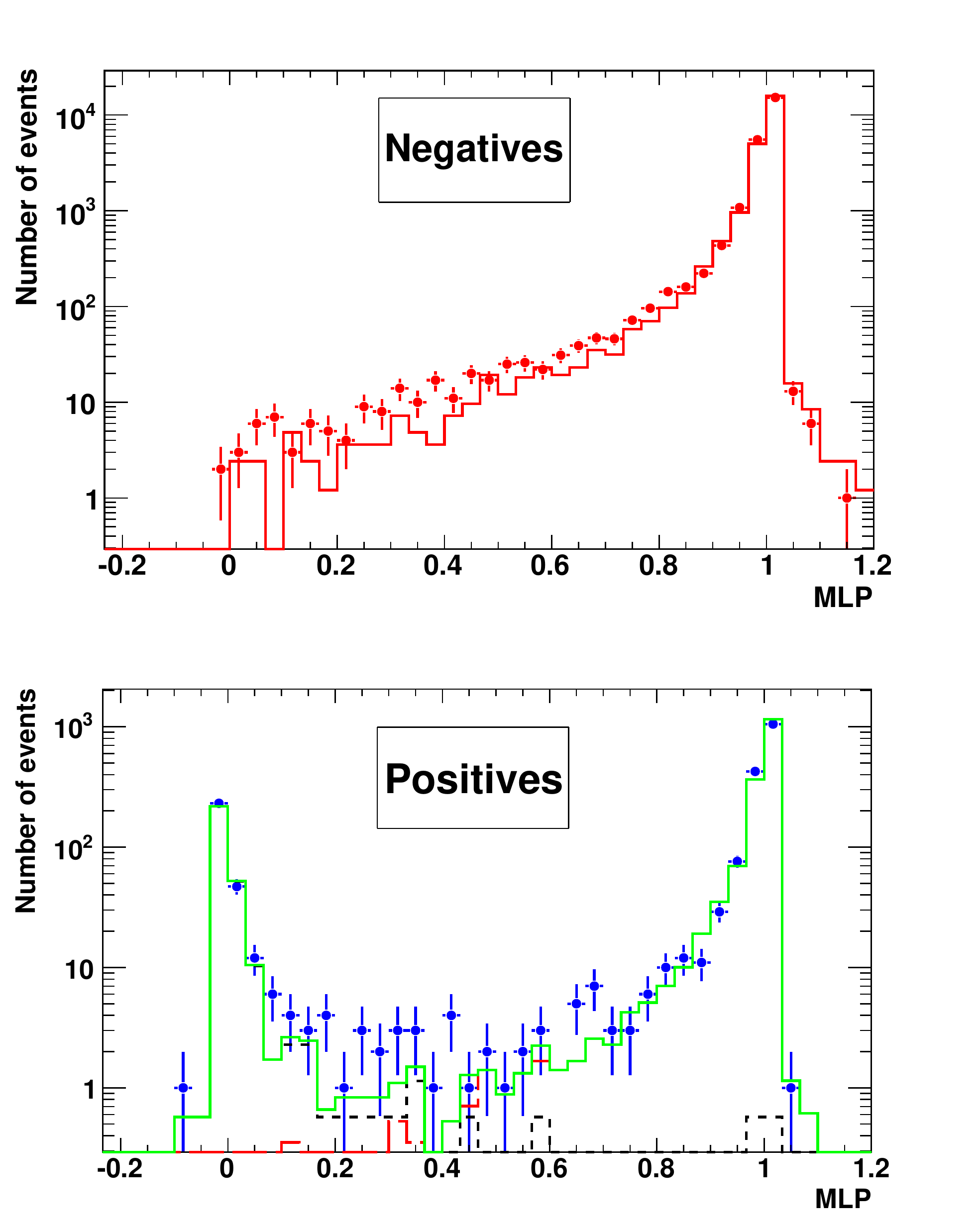}
  \caption{MLP output for negatives (top panel) and positives (bottom panel) particles.}
  \label{mlpout}
 \end{figure}
Figure \ref{mlpout} refers to the rigidity bin 2.1 to 2.4 GV and cutoff bin from zero up to 0.75 GV. In the top panel the MLP output distribution for selected 
negative particles (points) fitted using the shape of the distribution obtained from simulated electrons (solid line) is shown. The distribution for positive particles is displayed 
in the bottom panel. Events in the peak on the left correspond to the proton residual contamination, while the peak on the right is made of positron events. The distribution is 
fitted using the sum (solid line) of the simulated electrons (dashed line) and protons (dotted line) MLP distributions where their normalization factors are the free parameters. 
The normalization parameter obtained for the electron distribution represents the number of positron measurement. The chi square of the fit is generally very good, being the distributions 
dominated by the peaks with high statistics. The tails of the distributions coming from the simulations are less accurate, and indicate a non-perfect modelization of the 
instrument even if accurate Monte-Carlo tuning has been performed also with the help of test beam data. Anyhow, the disagreement between the simulated and real distribution tails gives a negligible 
systematic effect in the considered energy range.

For a given cutoff bin, the MLP spectral analisys is performed on all energy bins and a positron spectrum is obtained. A set of 15 continuous cutoff bins has been used in order to 
account for the spectral shapes at different latitudes. 
% figura output di neural network con spiegazioni
 \begin{figure}[t]
  \centering
  \includegraphics[width=0.4\textwidth]{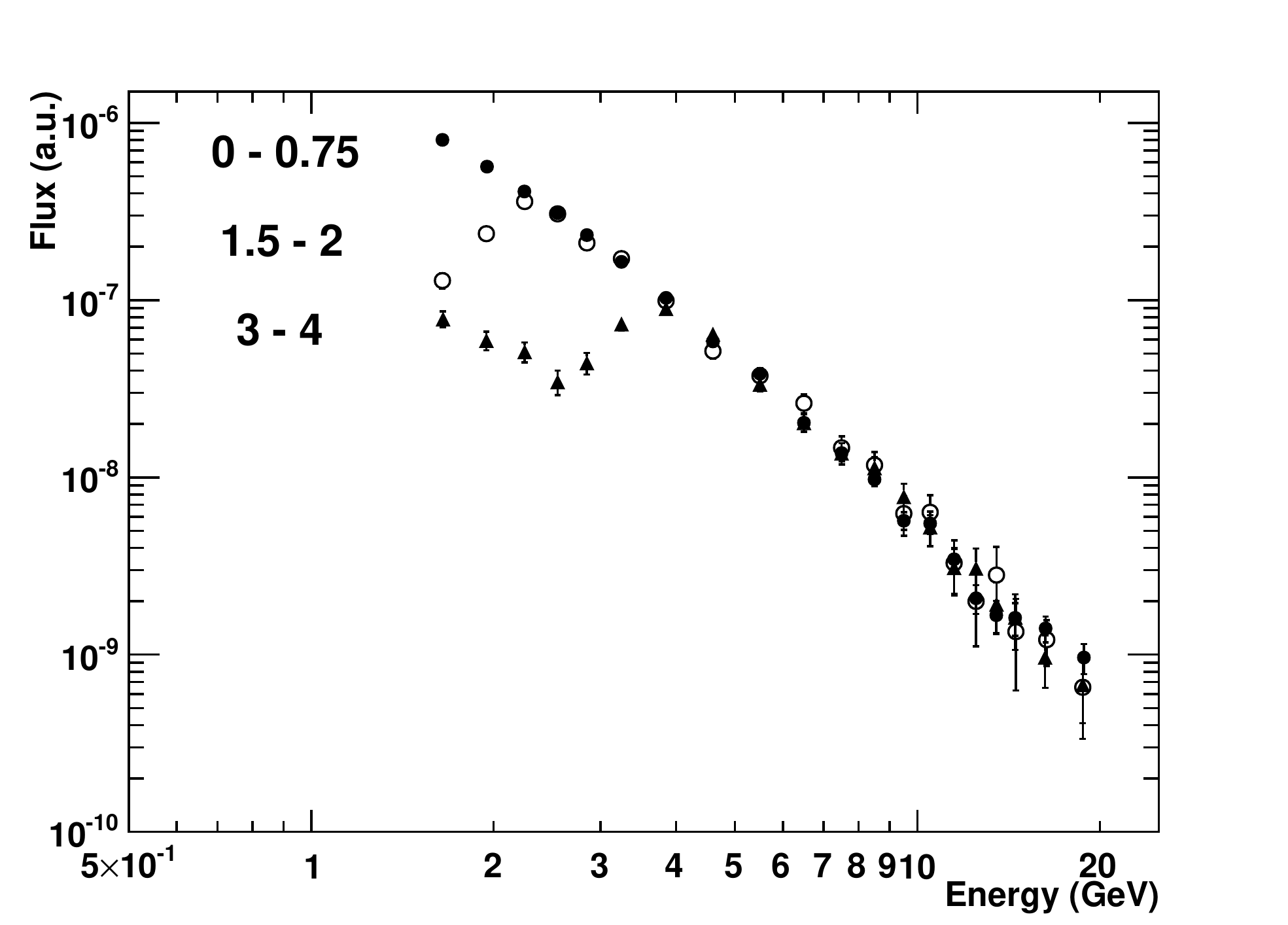}
  \caption{Positron fluxes prior to the unfolding procedure in three different cutoff bins.}
  \label{posifluxes}
 \end{figure}
Figure \ref{posifluxes} shows the positron fluxes prior to the unfolding procedure in three cutoff bins, from 0 to 0.75 GV (full circles), from 1.5 to 2 GV (open circles) and from 3 
to 4 GV (triangles). While in the first bin only galactic positrons are detected above 1.5 GV, in the last cutoff bin the decrease of the galactic spectrum can be clearly noticed 
together with the secondary re--entrant albedo flux below the cutoff.

Each spectrum is then unfolded using a Bayesan procedure \cite{dag95}. Only after the unfolding the cutoff selection, set to be 1.3 times the average vertical cutoff in the bin, is 
applied. Eventually, fluxes are summed to obtain the resulting spectrum in the whole energy range.

\subsection{Third momentum weighted average}
Particles with rigidity larger than 20 GV can always reach the Earth and no spectrum cutoff is observed at any satellite position. Hence no cutoff selection is required and the 
positron flux can be measured directly. However, with increasing energy the proton rejection becomes more difficult, both because of the lower energy resolution of the tracking system 
and because of the finite depth of the calorimeter (16 radiation length) which cannot fully contain anymore the whole electromagnetic showers. Observables used in the neural network 
algorithm gradually become less efficient in separating positrons from protons and the right tail of the proton distribution, figure \ref{mlpout}, becomes higher indicating an 
increased proton contamination in the positron sample. Using a conservative approach, we decided to change the positron selection method. As a first step all events of the 
preselected sample with MLP output less than 0.6 are discarded. The remaining events are electrons in the sample of negatives and are mostly positrons with a proton contamination component in 
the sample of positives. A set of four calorimetric observables, not included in the neural network analysis, is then used to build a third momentum variable event by event basis, that is 
$M^3 = \sum_{i=1}^4 X_i^3$ with $X_i = (x_{i} - \overline{x_i})/\sigma_{x_i} $ for the $i$-th observable. Variables are chosen in order to have a Gaussian distribution for positron 
events; mean and sigma of these distributions are measured using electron simulations and cross-checked using negative flight data. Due to its construction, the third momentum positron 
probability density function is symmetric around zero. On the contrary this distribution for simulated protons is proved, by the mean of simulations, to be asymmetric and centered to 
negative values. \cite{bia12} 
 \begin{figure}[t]
  \centering
  \includegraphics[width=0.4\textwidth]{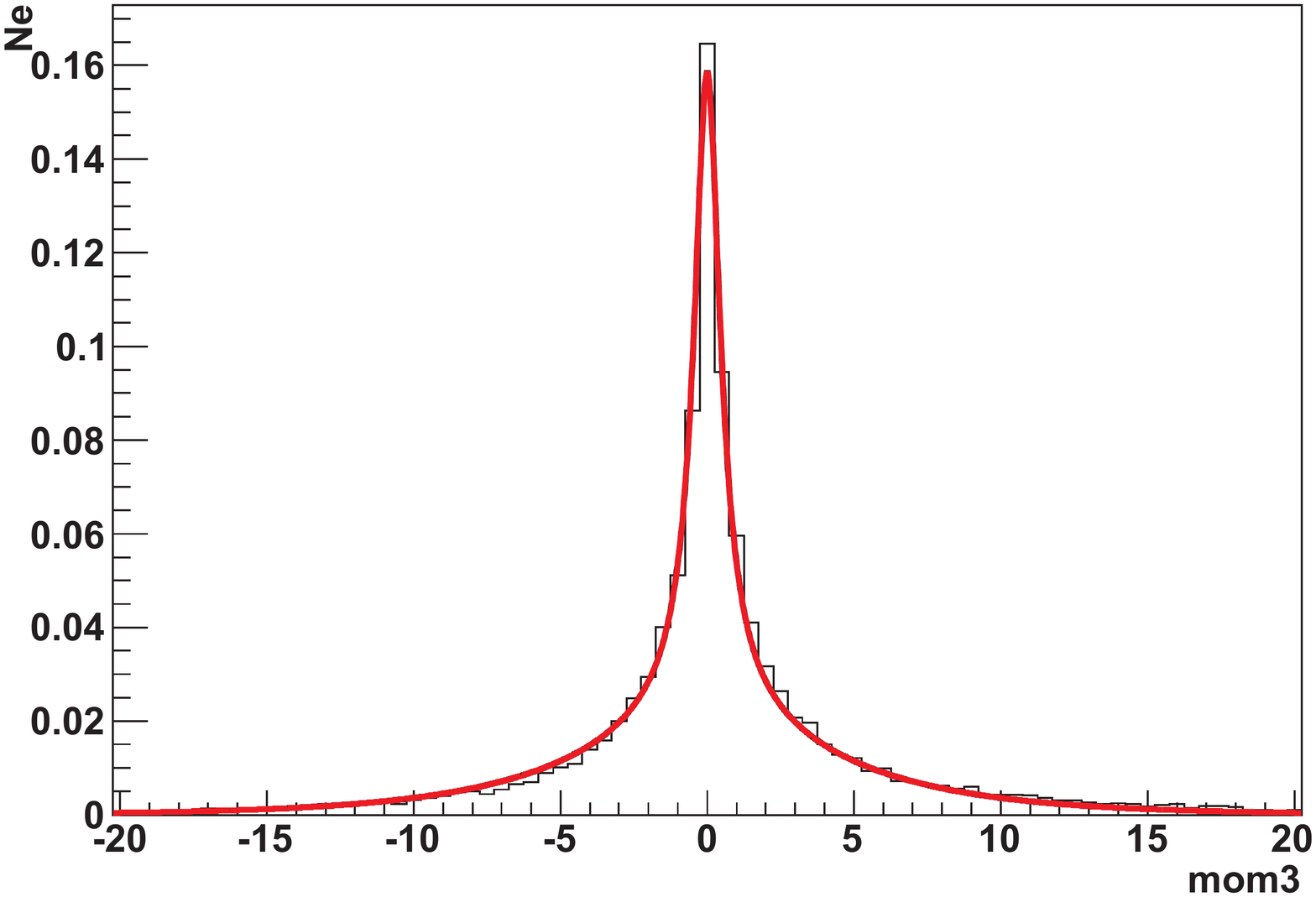}
  \includegraphics[width=0.4\textwidth]{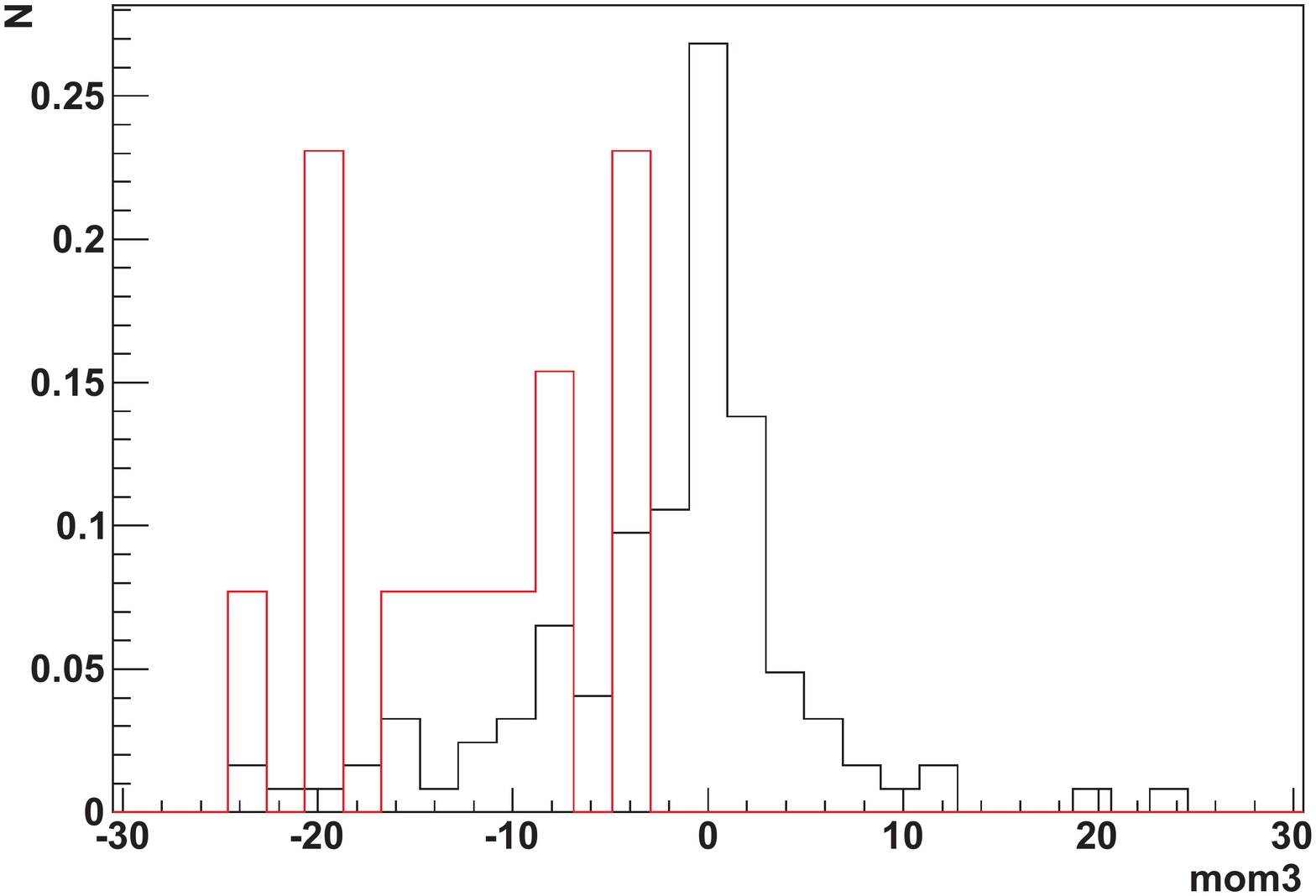}
  \caption{Top panel, normalized third momentum distribution for simulated electrons; the red line is a fit of the distribution. Bottom panel, third momentum distribution for positive data in the 
energy range 28--42 GV (black line) compared to the simulated protons one (red line).}
  \label{m3fit}
 \end{figure}
A fit of the third momentum distribution is performed on a sample of simulated electron and normalized to the unity, figure \ref{m3fit} top panel. Eventually a weighted average is used 
to determine the number of positron in the positive sample for each energy bin. In fact, the $M^3$ fitted function ($f(M^3)$) can be considered a probability distribution for 
electrons and positron events. A sample of $N_p$ positive events and a sample of $N_n$ negative events is selected, as described above; for each positive event, $p$, and 
each negative event, $n$, the corresponding value of $M^3(p/n)$ is calculated. Hence, the sum of the function value for each event can be calculated for positives and 
negatives events: $F_p=\sum_{i=1}^{N_p}f(M^3(p_{i}))$ and $F_n=\sum_{i=1}^{N_n}f(M^3(n_{i}))$. Assuming that all the negatives events $N_n$ are electrons (the antiproton and 
proton spillover contamination after the basic selection is negligible in this energy range) the number of positron events can be calculated as $(N_n/F_n) \cdotp F_p$.
Finally, a bootstrap procedure \cite{efr93} is applied to get the number of positrons for each energy bin.

\subsection{Lower limit estimation}
As energy increases, positron identification becomes more difficult, proton contamination becomes more significant and, moreover, simulation reproduces data with less accuracy. 
Furthermore, above 300 GeV charge sign confusion, i.e. electron spill-over contamination in the positron sample, is not negligible anymore. For these reasons, we decided to use the 
most reliable and worst possible scenario for proton contamination to determine if a positron signal can still be identified in the energy bin from 200 to 300 GeV.

By using real flight data we select events with absolute deflection less than about 0.001 GV, i.e. events with measured rigidity greater than the Maximum Detectable Rigidity 
(``over-MDR sample''). Since these events have an estimated error on the deflection greater than the deflection itself, the rigidity measured is meaningless and the sample can 
contain both positive and negative particles. Due to the harder spectral index of protons respect to the electrons, positrons and anti-protons one, the sample will probably contain 
mostly primary protons but the presence of electrons and/or positrons cannot be excluded. Moreover, this over-MDR sample represents the worst possible scenario of high energy protons 
emulating positrons and reconstructed in the wrong energy bin since the electromagnetic component of the hadronic shower is usually just a fraction of primary proton original 
energy. 

\begin{figure}[t]
  \centering              
  \includegraphics[width=0.4\textwidth]{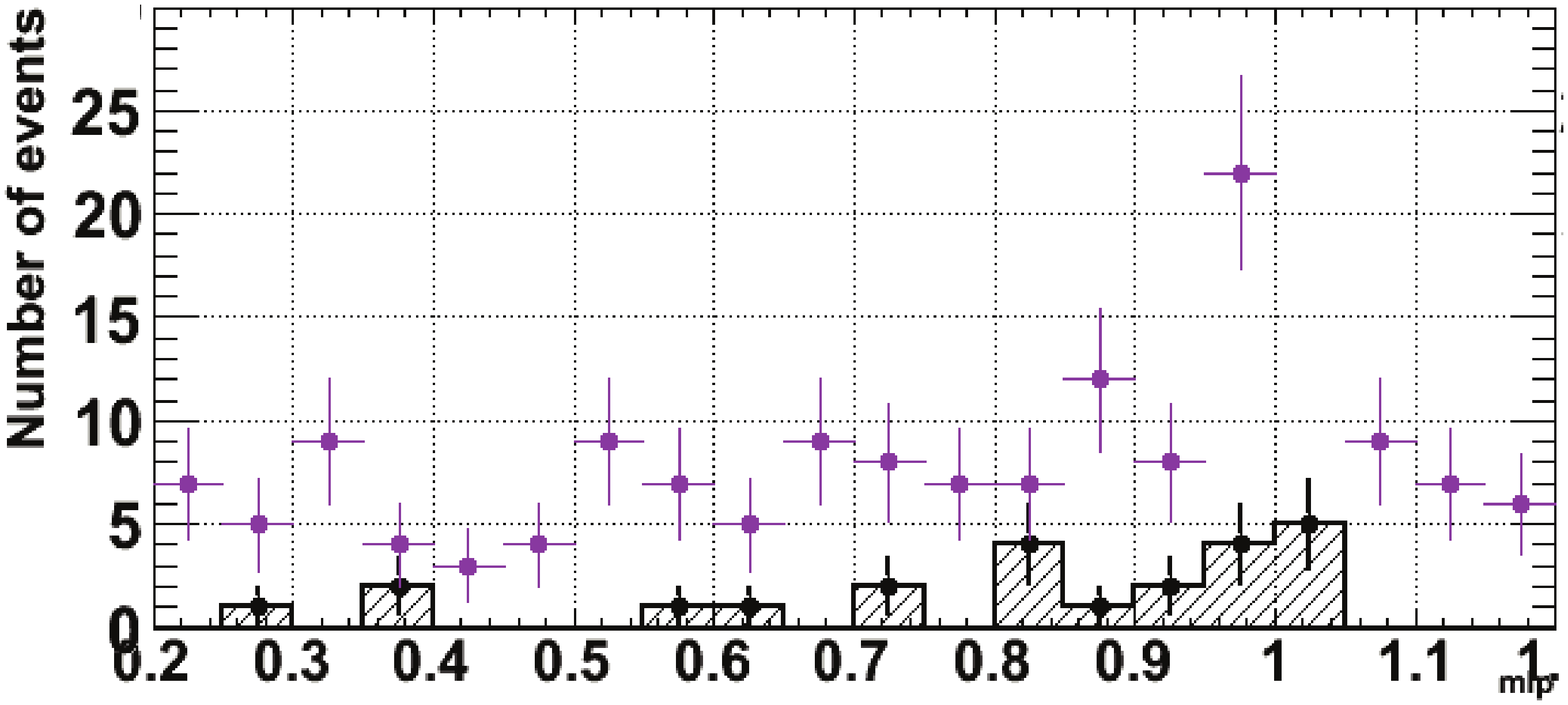}
  \caption{Neural network output, comparison between simulated protons (black line and circles) and over-MDR events distributions (purple circles).}
  \label{mlpzero}
 \end{figure}

The over-MDR sample is then re-processed as a sample of protons with energy between 200 to 300 GeV, where to each event a random energy 
in this interval is assigned following the flight-data real-proton energy distribution. The neural network algorithm is eventually applied to each event after the basic selection. 
%explained previously in this section. 
The resulting MLP output distribution is shown in figure \ref{mlpzero}.
Simulated protons distribution is shown in black, while over-MDR sample distribution is shown in purple. The two distributions are indeed different. It can be noticed how the tail of the over-MDR 
sample is much higher respect the simulated protons one and a small peak appears in correspondence to the positron peak position (MLP$\simeq $1). This peak could indicate either the presence of leptons 
in the over-MDR sample or that protons in this sample distribute in a different, and worst from the point of view of the proton rejection, respect the simulated ones in this energy range. Anyhow, 
similarly to what is done in the neural network spectral analysis, we fit the positives MLP output distribution with the sum of the simulated electrons and the over-MDR sample distribution shapes.

Interestingly, the peak and the tail found in the distribution of positives can be accounted only if to the over-MDR distribution a positron-like signal is added, even if a 
peak at MLP$\simeq$1 exists in the over-MDR sample. This result indicates that positrons are indeed present in the highest energy bin sample of positive events. However, by using the over-MDR 
distribution, we are over-estimating the proton contamination and only a lower limit can be determined. A bootstrap analysis is used to measure a lower limit with 90\% confidence 
level as a positron fraction and converted to a flux measurement using the PAMELA measured electron flux to determine the selection efficiencies.

\section{Efficiency estimation}
Positron selection efficiencies have been evaluated using a GEANT4 Monte-Carlo simulation. Efficiency validation was performed comparing, whenever possible, the simulated efficiency with the one 
measured using negatives flight data which consists mostly of electrons. 

\section{Results}
The same procedure used for positron was tested on the sample of negatives to obtain the electron flux and the resulting positron fraction. 
Results are in agreement with the previous PAMELA positron fraction published measurements \cite{psarticle,psarticle2}. 
The positron energy spectrum and the final positron fraction are presented in \cite{pssub}.
%The pure positron flux and the final positron fraction will be presented at the conference.

\section{Conclusions}
Positron flux and positron fraction are simultaneously measured by PAMELA. An increase in the positron flux measurement proves that the 
rise in the positron fraction observed with high significance for the first time by PAMELA and measured also by Fermi \cite{fermi} and AMS-02 \cite{ams13} suggests that a source of high energy 
positrons exist. More precise measurements will be needed to determine if the signal is coming from an astrophysical source or from decay and/or annihilation of dark matter particles.

\vspace*{0.5cm}
\footnotesize{{\bf Acknowledgment:}{We would like to acknowledge contributions and support from: Italian Space Agency (ASI), Deutsches Zentrum f\"ur Luft-- und Raumfahrt (DLR), The Swedish National Space Board, Swedish Research Council, Russian Space Agency (Roskosmos, RKA).  R. S. wishes to thank the TRIL program
 of the International Center of Theoretical Physics, Trieste, Italy that partly sponsored his
 activity.}}

\end{document}